\documentstyle[11pt]{article}
\begin{document}

\title{Non-commutative geometry of 4-dimensional quantum Hall droplet}

\author{Yi-Xin Chen$^1$ \thanks{Email:yxchen@zimp.zju.edu.cn} 
\hspace{.5mm}, Bo-Yu Hou$^2$ \thanks{Email:byhou@phy.nwu.edu.cn} 
\hspace{.5mm}, and Bo-Yuan Hou$^3$\\[.3cm]
$^1$Zhejiang Institute of Modern Physics, Zhejiang University,\\ 
             Hangzhou 310027, P. R.China,\\
$^2$Institute of Modern Physics, Northwestern University,\\
             Xi'an 710069, P. R. China\\
$^3$Graduate School, Chinese Academy of Science,\\
             Beijing 100039, P. R. China}

\date{}
\maketitle

\begin{abstract}

\indent

We develop the description of non-commutative geometry of the 
4-dimensional quantum Hall fluid's theory proposed recently by Zhang and 
Hu. The non-commutative structure of fuzzy $S^{4}$ , which is the base of the 
bundle $S^7$ obtained by the second Hopf fibration, i.e., $S^7/S^3 =S^4$, appears naturally in 
this theory. The fuzzy monopole harmonics, which are the essential 
elements in the non-commutative algebra of functions on $S^4$, are explicitly constructed and 
their obeying the matrix algebra is obtained. This matrix algebra is 
associative. We also propose a fusion scheme of the fuzzy monopole 
harmonics of the coupling system from those of the subsystems, and 
determine the fusion rule in such fusion scheme. By products, we provide 
some essential ingredients of the theory of $SO(5)$ angular momentum. In 
particular, the explicit expression of the coupling coefficients, in the 
theory of $SO(5)$ angular momentum, are given. We also discuss some 
possible applications of our results to the 4-dimensional quantum Hall 
system and the matrix brane construction in M-theory.

\vspace{.5cm}

\end{abstract}

\setcounter{equation}{0}

\section{Introduction}

\indent

The planar coordinates of quantum particles in the lowest Landau level of 
a constant magnetic field provide a well-known and natural realization of 
non-commutative space \cite{Dunne}.The physics of electrons in the lowest 
Landau level exhibits many fascinating properties. In particular, when the 
electron density lies in at certain rational fractions of the density 
corresponding to a fully filled lowest Landau level, the electrons 
condensed into special incompressible fluid states whose excitations 
exhibit such unusual phenomena as fractional charge and fractional 
statistics. For the filling fractions ${1 \over m}$, the physics of these 
states is accurately described by certain wave functions proposed by 
Laughlin \cite{Laughlin}, and more general wave functions may be used to 
describing the various types of excitations about the Laughlin states.

There has recently appeared an interesting connection between quantum 
Hall effect and non-commutative field theory. In particular, Susskind 
\cite{Susskind} proposed that non-commutative Chern-Simons theory on the 
plane may provide a description of the (fractional) quantum Hall fluid, 
and specifically of the Laughlin states. Susskind's non-commutative 
Chern-Simons theory on the plane describes a spatially infinite quantum 
Hall system. It, i.e., does the Laughlin states at filling fractions $\nu$ 
for a system of an infinite number of electrons confined in the lowest 
Landau level. The fields of this theory are infinite matrices which act on 
an infinite Hilbert space, appropriate to account for an infinite number of 
electrons. Subsequently, Polychronakos \cite{Polychronakos} proposed a 
matrix regularized version of Susskind's non-commutative Chern-Simons 
theory in an effort to describe finite systems with a finite number of 
electrons in the limited spatial extent. This matrix model was shown to 
reproduce the basic properties of the quantum Hall droplets and two special 
types of excitations of them. The first type of excitations is arbitrary 
area-preserving boundary excitations of the droplet. The another type of 
excitations are the analogs of quasi-particle and quasi-hole states. These 
quasi-particle and quasi-hole states can be regarded as non-perturbative 
boundary excitations of the droplet. Furthermore, it was shown that there 
exists a complete minimal basis of exact wave functions for the matrix 
regularized version of non-commutative Chern-Simons theory at arbitrary 
level $\nu^{-1}$ and rank $N$, and that those are one to one 
correspondence with Laughlin wave functions describing excitations of a 
quantum Hall droplet composed of $N$ electrons at filling fraction $\nu$ 
\cite{Hellerman}. It is believed that the matrix regularized version of 
non-commutative Chern-Simons theory is precisely equivalent to the theory 
of composite fermions in the lowest Landau level, and should provide an 
accurate description of fractional quantum Hall state. However, it does 
appear an interesting conclusion that they are agreement on the long 
distance behavior, but the short distance behavior is different 
\cite{Karabali}.

In the matrix regularized version of non-commutative Chern-Simons theory, 
a confining harmonic potential must be added to the action of this matrix 
model to keep the particles near the origin. In fact, there has been a 
translationally invariant version of Laughlin quantum Hall fluid for the 
two-dimensional electron gas in which it is not necessary to add any 
confining potential. Such model is Haldane' that of fractional quantum 
Hall effect based on the spherical geometry \cite{Haldane}. Haldane's 
model is set up by a two-dimensional electron gas of $N$ particles on a 
spherical surface in radial monopole magnetic field. A Dirac's monopole is 
at the center of two-dimensional sphere. This compact sphere space can be 
mapped to the flat Euclidean space by standard stereographical mapping. In 
fixed limit, the connection between this model and non-commutative 
Chern-Simons theory can be exhibited clearly. Exactly, the non-commutative 
property of particle's coordinates in Haldane's model should be described 
in terms of fuzzy two-sphere. 

In this paper, we do not plan to discuss 
such description of Haldane's model in detail, but want to exhibit the 
character of non-commutative geometry of 4-dimensional generalization of 
Haldane's model, proposed recently by Zhang and Hu \cite{Zhang}. The 
4-dimensional generalization of the quantum Hall system is composed of 
many particles moving in four dimensional space under a $SU(2)$ gauge 
field. Instead of the two-sphere geometry in Haldane's model, Zhang and Hu 
considered particles on a four-sphere surface in radial Yang's $SU(2)$ 
monopole gauge field \cite{Yang1}, which replaces the Dirac's monopole 
field of  Haldane's model. This Yang's  $SU(2)$ monopole gauge potential 
defined on four-sphere can be transformed to the instanton potential of the 
$SU(2)$ Yang-Mills theory \cite{Belavin} upon a conformal transformation 
from four-sphere to the 4-dimensional Euclidean space. Zhang and Hu had 
shown that at appropriate integer and fractional filling fractions the 
generalization of system forms an incompressible quantum fluid. They 
\cite{Hu} also investigated collective excitations at the boundary of the 
4-dimensional quantum Hall droplet proposed by them. In their discussion, 
an non-commutative algebraic relation between the coordinates of particle 
moving on four-sphere plays the key role. According to our understanding 
about Haldane's and Zhang et al works, we think that fuzzy sphere 
structures in their models is the geometrical origin of non-commutative 
algebraic relations of the particle coordinates. We shall clarify this 
idea in this paper.

Non-commutative spheres have found a variety of physical applications 
\cite{Madore,Grosse,de Wit,Grosse1,Connes,Bonechi,Konechny}. The 
description of fuzzy two-sphere \cite{Madore} was discovered in early attempts to 
quantize the super-membrane\cite{de Wit}. The fuzzy four-sphere appeared in 
\cite{Grosse,Grosse1}. The connection of non-commutative second 
Hopf bundle with the fuzzy four-sphere has been investigated \cite{Bonechi} from 
quantum group. The fuzzy four-sphere was used \cite{Castellino} in the context of the 
matrix theory of BFSS \cite{Banks} to described time-dependent 4-brane solutions constructed 
from zero-brane degrees of freedom. Furthermore, the non-commutative descriptions of 
spheres also arise in various contexts in the physics of D-branes. The 
descriptions of them, e.g., were used to exhibit the non-commutative 
properties and dielectric effects of D-branes \cite{Myers}. Recently, Ho 
and Ramgoolam \cite{Ramgoolam,Ho} had studied the matrix descriptions of 
higher dimensional fuzzy spherical branes in the matrix theory. They have 
found that the finite matrix algebras associated with the various fuzzy 
spheres have a natural basis which falls in correspondence with tensor 
constructions of irreducible representations of the corresponding 
orthogonal groups. In their formalism, they gave the connection between 
various fuzzy spheres and matrix algebras by introducing a projection from 
matrix algebra to fuzzy spherical harmonics. Their fuzzy spheres obey 
non-associative algebras because of the non-associativity induced by the 
projected multiplication. The complication of projection makes their 
constructions of fuzzy spherical harmonics formal.

The goal of this paper is to explore the character of non-commutative 
geometry of 4-dimensional quantum Hall system proposed recently by Zhang and Hu. 
Recently, Fabinger \cite{Fabinger} had pointed that there exists a connection of the fuzzy 
$S^4$ with Zhang and Hu's quantum Hall model of $S^4$. The string theory and brane matrix theory 
related with such non-commutative structure of fuzzy $S^4$ are discussed by \cite{Fabinger,Karabali1,Kimura}.
However, the structure of non-commutative algebra of functions on $S^4$ is not still clear. It is 
known that the key idea of non-commutative geometry is in replacement of commutative 
algebra of functions on a smooth manifold by a non-commutative deformation of it \cite{Konechny}.
We shall explore the structure the character of non-commutative 
geometry of 4-dimensional quantum Hall system to find non-commutative algebra of functions on $S^4$.

We should emphasize that the non-commutative structure of fuzzy $S^4$ of Zhang and Hu's quantum 
Hall model is different with those commonly considered by people. In fact, the $S^4$ of Zhang and Hu's 
quantum Hall model is the base of the Hopf bundle $S^7$ obtained 
by the second Hopf fibration, i.e., $S^7/S^3 =S^4$, from the connection between the second Hopf
map and Yang's $SU(2)$ \cite{Minami,Demler}. Equivalently, $\frac {SO(5)}{SU(2)\times SU(2)} =\frac {S^7} 
{SU(2)=S^3} =S^4 $ since $S^7 =\frac{SO(5)}{SU(2)}$. The bundle $S^7 =\frac {SO(5)}{SU(2)}$ can be parametered
$\theta, \alpha, \beta, \gamma, \alpha_I, \beta_I, \gamma_I$ (see section 2 in detail). By further smearing
out the common $U(1)$ gauge symmetry parametered by $\gamma_I$ , we can obtain the bundle $\frac {SO(5)}{SU(2)\times U(1)}
=\frac {SO(5)}{U(2)} =S^6 $ and its fibre $\frac {SU(2)}{U(1)} =\frac {S^3}{S^1} =S^2 $. Furthermore, we can find the sections
of the bundle $S^7$ or $S^6$ by solving the eigenfunctions of Zhang and Hu's model. The eigenfunctions of the LLL consist of
the space on which non-commutative algebra of functions on $S^4$ act.
   
Since at specialfilling factors, the quantum disordered ground 
state of 4-dimensional quantum Hall effect is separated from all excited 
states by a finite energy gap, the lowest energy excitations are 
quasi-particle or quasi-hole excitations near the lowest Landau level. The 
quantum disordered ground state of 4-dimensional quantum Hall effect is 
the state composed coupling by particles lying in the lowest Landau level 
state. At appropriate integer and fractional filling fractions, the system 
forms an incompressible quantum liquid, which is called as a 4-dimensional 
quantum Hall droplet. In fact, the spherical harmonic operators for fuzzy 
four-sphere are related with quasi-particle's or quasi-hole's creators of 
4-dimensional quantum Hall effect. These operators is composed of a 
complete set of the matrices with the fixed dimensionality. Focusing on 
the space of particle's lowest Landau level state, we shall construct 
explicitly these fuzzy spherical harmonics, also called the fuzzy 
monopole harmonics, and discuss the nontrivial algebraic relation between 
them. Furthermore, we shall clarify the physical implications of them.

This paper is organized as follows. Section two introduces the 
4-dimensional quantum Hall model proposed by Zhang and Hu, and analyzes 
the property of Hilbert space and the symmetrical structures of this 
4-dimensional quantum Hall system. We shall emphasize the intrinsic 
properties of the Yang's $SU(2)$ monopole included in this system, and 
give the explicit forms of normalized wave functions of this system, which is 
given by (8) in our paper. The wave functions corresponding to the irreducible representation 
$R=[r_1 = I,r_2 = I]$ of $SO(5)$ are those in the LLL. They are the sections of 
the bundle $S^7 =SO(5)/SU(2)$ over $S^4$ with fibre $SU(2)=S^3$ parametered 
by $\alpha_I, \beta_I, \gamma_I$. These degeneracy wave functions consist of the 
LLL Hilbert space which the non-commutative algebra of functions on $S^4$ acts on. The 
elements of this algebra are constructed in the following section. Section 
three describes the elements of non-commutative algebra of functions on $S^4$ which is 
related with the fuzzy four-sphere from the geometrical and symmetrical structures of 
4-dimensional quantum Hall droplet. We find the matrix forms and the symbols of the 
elements, given by (15) and (13) respectively. We give the explicit construction of 
complete set of this matrix algebra, determined by (19). In section four, we find the system 
of algebraic equations satisfied by the generators of the matrix algebra. The results are given 
by the equations (30) and (31). For the matrix forms of the elements, this non-commutative algebra
should be understood as the algebraic relation of matrix mulitiplication, and for the symbols of 
the elements, it should be done as that of Moyal product. It can be seen from these results and 
the complete set (19) that the non-commutative algebra is closed. The associativity of this algebra is 
shown by the relation (33). Furthermore, a fusion scheme of the fuzzy monopole harmonics of the 
coupling system from those of the subsystems, and its fusion rule are established in this 
section, which are given by the relations (40) and (41). Section five includes discussions about the physical 
interpretations of the results and remarks on some physical applications 
of them.

\section{The Hilbert space of 4-dimensional quantum Hall system}

The 4-dimensional quantum Hall system is composed of many particles moving 
in four dimensional space under a $SU(2)$ gauge field. The Hamiltonian of 
a single particle moving on four-sphere $S^{4}$ is read as
\begin{equation}
{\cal {H}}=\frac{\hbar^{2}}{2MR^{2}}\sum_{a<b}\Lambda_{ab}^{2}
\end{equation}
where $M$ is the inertia mass and $R$ the radius of $S^{4}$. The symmetry 
group of $S^{4}$ is $SO(5)$. Because the particle is coupling with a 
$SU(2)$ gauge field $A_{a}$, $\Lambda_{a b}$ in Eq.(1) is the dynamical 
angular momentum given by $\Lambda_{a b} = -i(x_a D_b - x_b D_a)$. From 
the covariant derivative $D_a=\partial_a + A_a$, one can calculate the 
gauge field strength from the definition $f_{a b} = [D_a, D_b]$. 
$\Lambda_{a b}$ does not satisfy the commutation relations of $SO(5)$ 
generators. Similar to in Dirac's monopole field, the angular momentum of 
a particle in Yang's $SU(2)$ monopole field can be defined as $L_{a b} = 
\Lambda_{a b} - i f_{a b}$, which indeed obey the $SO(5)$ commutation 
relations. Yang\cite{Yang} proved that $L_{a b}$ can generate all $SO(5)$ 
irreducible representations.

In general, the representations of $SO(5)$ can be put in one-to-one 
correspondence with Young diagrams, labelled by the row lengths $[r_1, 
r_2]$, which obey the constraints $0 \leq r_2 \leq r_1$. For such a 
representation, the eigenvalue of Casimir operator is given by $A(r_1, 
r_2) = \sum_{a < b} L_{a b}^2 = r_1^2 + r_2^2 + 3 r_1 + r_2$, and its 
dimensionality is
\begin{equation}
D(r_1, r_2) = \frac{1}{6}(1 + r_1 - r_2)(1 + 2 r_2)(2 + r_1 + r_2)(3 + 2 
r_1).
\end{equation}
The $SU(2)$ gauge field is valued in the $SU(2)$ Lie algebra 
$[I_i, I_j]=i \varepsilon_{i j k} I_k$. The value of this $SU(2)$ Casimir 
operator $\sum_{i} I_i^2 = I (I + 1)$ specifies the dimension of the 
$SU(2)$ representation in the monopole potential. $I$ is an important 
parameter of $L_{a b}$ generating all $SO(5)$ irreducible representations 
and the Hamiltonian Eq.(1). In fact, for a given $I$, if one deals with 
the eigenvalues and eigenfunctions the operator of angular momentum 
$\sum_{a < b} L_{a b}^2$, it can be found that the $SO(5)$ irreducible 
representations, which the eigenfunctions called by Yang \cite{Yang} as 
$SU(2)$ monopole harmonics belong to, are restricted. Such $SO(5)$ 
irreducible representations are labelled by the integers $[r_1, r_2 = I]$, 
and $r_1 \geq I$. Based on the expressions of $SU(2)$ monopole potentials 
given by Yang \cite{Yang}, or by Zhang and Hu \cite{Zhang}, one can show 
that $\sum_{a < b} \Lambda_{a b}^2 = \sum_{a < b} L_{a b}^2 - \sum_{i} 
I_i^2$ by straightforwardly evaluating. This implies that the eigenvalues 
and eigenfunctions of the Hamiltonian Eq.(1) can be read off from those of 
the operator $\sum_{a < b} L_{a b}^2$. Hence, for a given $I$, the energy 
eigenvalues of the Hamiltonian Eq.(1) are read as
\begin{equation}
E_{[r_1, r_2 = I]} = \frac{\hbar^2}{2M R^2}[A(r_1, r_2 = I) - 2 I(I + 1)].
\end{equation}
The degeneracy of energy level is given by the dimensionality of the 
corresponding irreducible representation $D(r_1, r_2 = I)$.

The ground state of the Hamiltonian Eq.(1) plays a key role in the 
procedure of construction of many-body wave function and the discussion of 
incompressibility of 4-dimensional quantum Hall system. This ground state, 
also called the lowest Landau level (LLL) state, is described by the least 
admissble irreducible representation of $SO(5)$, i.e., labelled by $[r_1 = 
I, r_2 = I]$ for a given $I$. The LLL state is $D(r_1 = I, r_2 = I) = 
\frac{1}{6}(2 I + 1)(2 I + 2)(2 I + 3)$ fold degenerate, and its energy 
eigenvalue is $\frac{\hbar^2}{2M R^2} 2 I$. Zhang and Hu \cite{Zhang} 
found the explicit form of the ground state wave function in the spinor 
coordinates. This wave function is read as
\begin{equation}
\langle x_a, {\bf n}_i | m_1, m_2, m_3, m_4 \rangle = \sqrt{\frac{p!}{m_1! 
m_2! m_3! m_4!}} \Psi_1^{m_1} \Psi_2^{m_2} \Psi_3^{m_3} \Psi_4^{m_4},
\end{equation}
with integers $m_1 + m_2 + m_3 + m_4 = p = 2 I$. The orbital coordinate 
$x_a$, which is defined by the coordinate point of the 4-dimensional 
sphere $X_a = R x_a$, is related with the spinor coordinates $\Psi_\alpha$ 
with $\alpha=1, 2, 3, 4$ by the relations $x_a = \bar{\Psi} \Gamma_a \Psi$ 
and $\sum_{\alpha} \bar{\Psi}_\alpha \Psi_\alpha = 1$. The five $4 \times 
4$ Dirac matrices $\Gamma_a$ with  $a=1, 2, 3, 4, 5$ satisfy the Clifford 
algebra $\{\Gamma_a, \Gamma_b \} = 2 \delta_{a b}$. The isospin 
coordinates ${\bf n}_i = \bar{u}{\bf \sigma}_i u$ with $i=1, 2, 3$ are 
given by an arbitrary two-component complex spinor $(u_1, u_2)$ satisfying 
$\sum_{\sigma} \bar{u}_\sigma u_\sigma = 1$. Zhang and Hu gave the 
explicit solution of the spinor coordinate with respect to the orbital 
coordinate as following
\begin{equation}
\left ( \begin{array} {l}
\Psi_1 \\
\Psi_2 \end{array} \right ) = \sqrt{\frac{1 + x_5}{2}} \left ( 
\begin{array} {l}
u_1 \\
u_2 \end{array} \right ),\quad
\left ( \begin{array} {l}
\Psi_3 \\
\Psi_4 \end{array} \right ) = \sqrt{\frac{1}{2(1+x_5)}} (x_4 - i x_i 
\sigma_i) \left ( \begin{array} {l}
u_1 \\
u_2 \end{array} \right ).
\end{equation}
By computing the geometric connection, one can get a non-Abelian gauge 
potential $A_a$, which is just the $SU(2)$ gauge potential of a Yang 
monopole defined on 4-dimensional sphere $S^{4}$ \cite{Yang1, Zhang}. 
Since we do not need the explicit form of it here, we do not write out 
that of it.

The description of the 4-dimensional quantum Hall liquid involves the 
quantum many-body problem of $N$ particle's moving on the 4-dimensional 
sphere $S^{4}$ in the Yang's $SU(2)$ monopole field lying in center of the 
sphere $S^{4}$. The wave functions of many particles can be constructed by 
the nontrivial product of the single particle wavefunctions, among which 
every single particle wave function is given by the LLL wavefunction 
Eq.(4). In the case of integer filling, the many-particle wave function is 
simply the Slater determent composed of $N$ single-particle wave 
functions. For the fractional filling fractions, the many particle wave 
function cannot be expressed as the Laughlin form of a single product. But 
the amplitude of the many-particle wave function can also be interpreted 
as the Boltzmann weight for a classical fluid. One can see that it 
describes an incompressible liquid by means of plasma analogy. Therefore, 
at the integer or fractional filling fractions, the 4-dimensional system 
of the generalizing quantum Hall effect forms an incompressible 
quantum liquid \cite{Zhang}. We shall call this 4-dimensional system as a 
4-dimensional quantum Hall droplet.

The space of the degenerate states in the LLL is very important not only 
for the description of the 4-dimensional quantum Hall droplet but also for 
that of edge excitations and quasi-particle or quasi-hole excitations of 
the droplet. In fact, this space of the degenerate states is the space 
which we shall construct the matrix algebra acting on in the following 
section. In order to construct the complete set of matrix algebra of fuzzy 
$S^{4}$, we need to know the explicit forms of the wave functions 
associated with all irreducible representations $[r_1, r_2 = I]$ of 
$SO(5)$. Although Yang had found the wave functions for all the $[r_1, r_2 
= I]$ states, the form of his parameterizing the four sphere $S^{4}$ is 
not convenient for our purpose. Following Hu and Zhang\cite{Hu}, we can 
parameterize the four sphere $S^{4}$ by the following coordinate system
\begin{eqnarray}
x_1 & = & \sin \theta \sin \frac{\beta}{2} \sin(\alpha - \gamma), 
\nonumber\\
x_2 & = & - \sin \theta \sin \frac{\beta}{2} \cos(\alpha - \gamma), 
\nonumber\\
x_3 & = & - \sin \theta \cos \frac{\beta}{2} \sin(\alpha + \gamma), 
\nonumber\\
x_4 & = & \sin \theta \cos \frac{\beta}{2} \cos(\alpha + \gamma), 
\nonumber\\
x_5 & = & \cos \theta,
\end{eqnarray}
where $\theta, \beta \in[0, \pi)$ and $\alpha, \gamma \in[0, 2\pi)$. The 
direction of the isospin is specified by $\alpha_I, \beta_I$ and 
$\gamma_I$.

As the above explanation, we can get the eigenfunctions of the Hamiltonian 
Eq.(1) from the eigenfunctions of the operator $\sum_{a < b}L_{a b}^2$. 
The angular momentum operators $L_{a b}$ consists of an orbital part 
$L_{\mu \nu}^{(0)} = - i (x_\mu \partial_\nu - x_\nu \partial_\mu)$ with 
$\mu, \nu=1, 2, 3, 4$ and an isospin part involving the $SU(2)$ monopole 
field. The angular momentum operators $L_{\mu \nu}^{(0)}$ generate the 
rotation in the subspace $(x_1, x_2, x_3, x_4)$, and satisfy the 
commutation relations of $SO(4)$ generators. They can be decomposed into 
two $SU(2)$ algebras: ${\hat J}_{1 i}^{(0)} = \frac{1}{2} (\frac{1}{2} 
\varepsilon_{i j k} L_{j k}^{(0)} + L_{4 i})$ and ${\hat J}_{2 
i}^{(0)} = \frac{1}{2} (\frac{1}{2} \varepsilon_{i j k} L_{j k}^{(0)} - 
L_{4 i})$. They satisfy the identity of operator $\sum_{i} {\hat J}_{1 
i}^{(0) 2} = \sum_{i} {\hat J}_{2 i}^{(0) 2}$. Therefore, if one would use 
the operators ${\hat J}_{1 i}^{(0)}$ and ${\hat J}_{2 i}^{(0)}$ to 
generate the $SO(4)$ blocks of the $SO(5)$ irreducible representations, he 
can not obtain all irreducible representations of $SO(5)$. However, 
because of the coupling to the Yang's $SU(2)$ monopole potential, these 
orbital $SO(4)$ generators are modified into $L_{\mu \nu}$, which are 
decomposed into ${\hat J}_{1 i} = {\hat J}_{1 i}^{(0)}$ and ${\hat J}_{2 
i} = {\hat J}_{2 i}^{(0)} + {\hat I}_i$. Indeed, the $SO(4)$ generators 
${\hat J}_{1 i}$ and ${\hat J}_{2 i}$ can be used to generate the $SO(4)$ 
block states of all $SO(5)$ irreducible representations \cite{Yang}. Such 
$SO(4)$ block states can be labelled by the $SO(4)$ quantum numbers 
$J \equiv \left ( \begin{array} {ll}
j_1 & j_2 \\
j_{1 z} & j_{2 z} \end{array} \right )$, where $\sum_{i} {\hat J}_{1 i}^2 
=j_1(j_1 + 1)$ and $\sum_{i} {\hat J}_{2 i}^2 = j_2(j_2 + 1)$. The $j_{1 
z}$ and $j_{2 z}$ are the magnetic quantum numbers of two $SU(2)$ 
algebras. 

Applying the $SO(4)$ operators to the quantum states described by the 
Hamiltonian Eq.(1) in the irreducible representations of $SO(5)$, one can 
see that the complete set of quantum observables of the system is composed 
of the operators $\sum_{a < b} L_{a b}^2, {\hat J}_1^2, {\hat J}_2^2, 
\hat{J}_{1 z}$ and ${\hat J}_{2 z}$. Thus, there exist the simultaneous 
eigenfunctions of those operators, which are just the wave functions of 
energy eigenvalue being $E_{[r_1, r_2 = I]}$. Noticing that the isospin 
operators ${\hat I}_i$ are coupling with the operator $\hat{J}_{2 
i}^{(0)}$ into the angular momentum operators ${\hat J}_{2 i}$, we should 
also introduce the quantum number labelling the isospin parameter $I$, 
which is written as $I^{(0)} \equiv \left ( \begin{array} {ll}
0 & I \\
0 & I_z \end{array} \right )$. The parameters of group $(\theta, \alpha, 
\beta, \gamma, \alpha_I, \beta_I, \gamma_I)$ are abbreviated to 
$(\Omega)$. The wave functions corresponding to the irreducible 
representation $R\equiv[r_1, r_2=I]$ of $SO(5)$ are denoted as ${\cal 
D}_{J I^{(0)}}^{(R)}(\Omega)$. By using the equation of parameterizing 
$S^{4}$ and the parameters of the isospin direction, and following Yang 
\cite{Yang}, one can obtain the wave functions ${\cal D}_{J 
I^{(0)}}^{(R)}(\Omega)$ obeying the system of equations as following
\begin{eqnarray}
& & {\hat J}_1^2 {\cal D}_{J I^{(0)}}^{(R)}(\Omega) = j_1(j_1 + 1){\cal 
D}_{J I^{(0)}}^{(R)}(\Omega), \quad
{\hat J}_{1 z} {\cal D}_{J I^{(0)}}^{(R)}(\Omega) = j_{1 z} {\cal D}_{J 
I^{(0)}}^{(R)}(\Omega), \nonumber \\
& & {\hat I}^2 {\cal D}_{J I^{(0)}}^{(R)}(\Omega) = I(I + 1) {\cal D}_{J 
I^{(0)}}^{(R)}(\Omega), \nonumber \\
& & {\hat J}_2^2 {\cal D}_{J I^{(0)}}^{(R)}(\Omega) = j_2(j_2 + 1){\cal 
D}_{J I^{(0)}}^{(R)}(\Omega),\quad
{\hat J}_{2 z} {\cal D}_{J I^{(0)}}^{(R)}(\Omega) = j_{2 z}{\cal D}_{J 
I^{(0)}}^{(R)}(\Omega), \nonumber \\
& & \{ \frac{1}{\sin^3 \theta} \frac{\partial}{\partial\theta} (\sin^3 
\theta \frac{\partial}{\partial\theta}) - \frac{4 {\hat J}_1^2}{\sin^2 
\theta} - \frac{2(1 - \cos \theta)}{\sin^2 \theta}({\hat J}_2^2 - {\hat 
J}_1^2 - {\hat I}^2) \nonumber\\
& & - \frac{(1 - \cos \theta)(3 + \cos \theta)}{\sin^2 
\theta} {\hat I}^2 \} {\cal D}_{J I^{(0)}}^{(R)}(\Omega) 
= - A(r_1, r_2 = I) {\cal D}_{J I^{(0)}}^{(R)}(\Omega).
\end{eqnarray}

The first line and second line of the equations tell us that we can use two $SU(2)$ 
D-functions to realize the wave function with respect to the dependence of 
the group parameters $\alpha, \beta, \gamma$ and $\alpha_I, \beta_I, 
\gamma_I$. Furthermore, we also should consider the coupling relations 
${\hat J}_{2 i} = {\hat J}_{2 i}^{(0)} + {\hat I}_i$, ${\hat J}_{1 i} = 
{\hat J}_{1 i}^{(0)}$ and $\sum_{i} {\hat J}_{1 i}^{(0) 2} = \sum_{i} 
{\hat J}_{2 i}^{(0)2}$ to make the wave function obey the third line of 
the equations. The final line is the equation to determine the $\theta$ 
dependence of the wave function, which had been solved by Yang 
\cite{Yang}. Now, we can write the explicit solution form of the 
normalized wavefunction as
\begin{eqnarray}
{\cal D}_{J I^{(0)}}^{(R)}(\Omega) & = & \sqrt{\frac{2 r_1 + 3}{2}} (\sin 
\theta)^{-1} d^{(r_1 + 1)}_{j_1 - j_2, - j_1 - j_2 - 1}(\theta) 
\frac{\sqrt{(2 I + 1)(2 j_1 + 1)}}{8 \pi^2} \nonumber\\
& \times & \sum_{j_{2 z}^{(0)}, I_z^\prime} D_{j_{1 z} j_{2 
z}^{(0)}}^{j_1}(\alpha, \beta, \gamma) \langle j_2^{(0)} = j_1, j_{2 
z}^{(0)}; I, I_z^\prime | j_2, j_{2 z} \rangle D_{I_z^\prime 
I_z}^{(I)}(\alpha_I, \beta_I, \gamma_I),
\end{eqnarray}
where
\begin{eqnarray}
d^{(r_1 + 1)}_{j_1 - j_2, - j_1 - j_2 - 1}(\theta) & = & \sqrt{\frac{2^{- 
2(j_1 - j_2)} (r_1 + j_1 - j_2 + 1)! (r_1 - j_1 + j_2 + 1)!}{(r_1 + j_1 + 
j_2 + 2)! (r_1 - j_1 - j_2)!}} \nonumber\\
& \times & (1 - \cos \theta)^{\frac{2 j_1 + 1}{2}} (1 + \cos \theta)^{- 
\frac{2 j_2 + 1}{2}} P^{(2 j_1 + 1,- 2 j_2 - 1)}_{r_1 - j_1 + j_2 + 
1}(\cos \theta),
\end{eqnarray}
and $P^{(\alpha, \beta)}_n$ is the Jacobi polynomial. 

Although $d^{(r_1 + 1)}_{j_1 - j_2,- j_1 - j_2 - 1}(\theta)$ is the 
d-function of $SO(3)$ rotation group, it should be emphasized that 
here the quantum numbers of the subgroup $SO(4)$ of $SO(5)$ have replaced 
the usual magnetic quantum number of $SO(3)$. Practically, $\sqrt{\frac{2 r_1 + 3}{2}} (\sin 
\theta)^{-1} d^{(r_1 + 1)}_{j_1 - j_2, - j_1 - j_2 - 1}(\theta)$ is the
d-function of $SO(5)$ rotation group in the special case. The D-function 
$D_{j_{1 z} j_{2 z}^{(0)}}^{j_1}(\alpha, \beta, \gamma)$ is the standard $SU(2)$ 
representation matrix for the Euler angles $\alpha, \beta, \gamma$. They 
are the $SU(2)$ rotation matrix elements generated rotationally by the 
operators ${\hat J}_{1 i}$. Similarly, $D_{I_z^\prime I_z}^{(I)}(\alpha_I, 
\beta_I, \gamma_I)$ are those generated by the isospin operators ${\hat 
I}_i$. The coupling coefficients $\langle j_2^{(0)} = j_1, j_{2 z}^{(0)}; 
I, I_z^\prime | j_2, j_{2 z} \rangle$, i.e., the Clebsch-Gordon 
coefficients, show the coupling behavior of the $SU(2)$ angular momentums 
${\hat J}_{1 i}$, ${\hat I}_i$ and ${\hat J}_{2 i}$ by means of ${\hat 
J}_{2 i}^{(0)}$. 

The isospin direction can be normally specified by two angles $\alpha_I$ 
and $\beta_I$. However, the wave function ${\cal D}_{J 
I^{(0)}}^{(R)}(\Omega)$ depends on the Eular angles $\alpha_I$, $\beta_I$ 
and $\gamma_I$ of the isospin space. In fact, in the $({\hat I}^2, {\hat 
I}_z)$ picture, the $\gamma_I$ dependence of ${\cal D}_{J 
I^{(0)}}^{(R)}(\Omega)$ is given by the $U(1)$ phase factor $\exp \{-i I_z 
\gamma_I \}$, where $I_z$ is simply an $U(1)$ gauge index. Therefore, 
different values of $I_z$ correspond to the same physical state. One can 
smear the $\gamma_I$ dependence of the wave function by the gauge choice. 
Such gauge choice can be fixed by taking $I_z = I$ or $I_z = - I$. If such 
gauge choice is taken at every step in all calculations, we call this 
choice as taking the physical gauge. Analogous to doing usually in the 
field theory, there exists another gauge choice, which such gauge choice 
is taken at the end of calculations. The latter gauge choice are called as 
taking the covariant gauge. We shall take the covariant gauge in this 
paper, which this thick was also used in the reference \cite{Hu}. 

The wave functions ${\cal D}_{J I^{(0)}}^{(R)}(\Omega)$ are the $SU(2)$ 
monopole harmonics. Exactly, they are the spherical harmonics on the coset 
space $SO(5)/SU(2)$, which is locally isomorphic to the sphere 
$S^{4}\times S^{3}$. By smearing the $U(1)$ degree of freedom of the 
$\gamma_I$ dependence, i.e., taking the physical gauge, they can be viewed 
as the spherical harmonics on the coset space $SO(5)/U(2)$, which is 
locally isomorphic to the sphere $S^{4}\times S^{2}$. Globally, 
$SO(5)/U(2)$ is a bundle over the $S^{4}$ with fibre $S^{2}$. The wave 
functions are the cross sections in this nontrivial fibre bundle. This 
implies that there exists a stabilizer group of the wave function 
solutions $U(2)$, and the action of $SO(5)$ on the state generates a space 
of the wave function solutions which is the space of cross sections in 
$SO(5)/U(2)$. If we take the covariant gauge, the $SU(2)$ monopole 
harmonics should be regarded as the cross sections in the nontrivial fibre 
bundle $SO(5)/SU(2)$. Then, the stabilizer group of the wave function 
solutions is $SU(2)$. $SO(5)$ doing the state generates that of cross 
sections in $SO(5)/SU(2)$. The procedure of parameterizing the four sphere 
$S^{4}$ Eq.(6) and building up the isomorphic relation of the $SU(2)$ 
group manifold and the three sphere $S^{3}$ is just that of smearing the 
stabilizer subgroup $SU(2)$. Of course, if one want to take the physical 
gauge, he can smear the $U(2)$ by further smearing the $U(1)$ gauge 
subgroup. Such globally geometrical structure and symmetrical structure 
can guide us to develop some techniques which we need in this paper.

Let us introduce the state vector $\left | \begin{array} {l}
R \\
J \end{array} \right \rangle$, which belongs to the Hilbert space ${\cal 
H}^{(R)}$ composed of the $SU(2)$ monopole harmonics 
${\cal D}_{J I^{(0)}}^{(R)}(\Omega)$. Of course, 
$\left | \begin{array} {l}
R \\
I^{(0)} \end{array} \right \rangle$ is an element of the Hilbert space 
${\cal H}^{(R)}$. In general, taking a fixed vector in the Hilbert space, 
one can use the unitary irreducible representation of an arbitrary Lie 
group acting in the Hilbert space to produce the coherent state for this 
Lie group. Now, we are interesting to the coherent state corresponding to 
the coset space $SO(5)/SU(2)$. The $SU(2)$ is the isotropic subgroup of 
$SO(5)$ for the state $\left | \begin{array} {l}
R \\
I^{(0)} \end{array} \right \rangle$ since $SU(2)$ is the stabilizer group 
of the wave function solutions. If we use the unitary irreducible 
representations of $SO(5)$ acting on the state $\left | \begin{array} {l}  
R \\
I^{(0)} \end{array} \right \rangle$ to produce the coherent states, the 
coherent state vectors belonging to a left coset class of $SO(5)$ with 
respect to the subgroup $SU(2)$ differ only in a phase factor and so 
determine the same state. Consequently, the coherent state vectors depend 
only on the group parameters parameterizing the coset space $SO(5)/SU(2)$. 
Thus, we can now introduce the following coherent state vector
\begin{equation}
|\Omega , R \rangle = \sum_{J, I_z} {\cal D}_{J 
I^{(0)}}^{*(R)}(\Omega) \left | \begin{array} {l}
R \\
J 
\end{array} \right \rangle |I, I_z \rangle ,
\end{equation}
where the star stands for the complex conjugate. In order to realize the 
covariant gauge, we have added the isospin frame $ |I, I_z \rangle $ to the 
monopole harmonics ${\cal D}_{J I^{(0)}}^{*(R)}(\Omega)$. In fact, we can 
use the finite rotation ${\hat {\cal R}}(\Omega) = \exp \{i \alpha {\hat 
J}_{1 z} \} \exp \{i \beta {\hat J}_{1 y}\} \exp \{i \gamma {\hat J}_{1 z} 
\} \exp \{i \theta {\hat L}_{5 4}\} \exp \{i \alpha_I {\hat I}_z\} \exp 
\{i \beta_I {\hat I}_y \} \exp \{i \gamma_I {\hat I}_z \}$ acting on the 
state vector $\left | \begin{array} {l}
R \\
I^{(0)} \end{array} \right \rangle$ to realize the general finite rotation 
of $SO(5)$ acting on the state vector $\left | \begin{array} {l}
R \\  
I^{(0)} \end{array} \right \rangle$ up to a phase factor, and to generate 
the above coherent state vector. 

The wave functions in the coherent state picture are given by
\begin{equation}
\left \langle \Omega , R \left | \begin{array} {l}
R \\ 
J
\end{array} \right \rangle  = \sum_{I_z} \langle I, I_z |{\cal D}_{J 
I^{(0)}}^{(R)}(\Omega) \right. .
\end{equation}
The l.h.s. of Eq.(11) is not with the label $I$ since it naturely appears in
the label $R=[r_1 , r_2 =I]$, which corresponds to the $SU(2)$ monopole
harmonics. It should be emphasized that the above wave functions become the wave 
functions in the physical gauge only if one smears the isospin frame of 
them and projects back to the $\alpha_I$ and $\beta_I$ angles. In the 
sense of the finite rotation of $SO(5)$, the explicit forms of wave 
function solutions given here are the wave function solutions of the 
4-dimensional spherically symmetrical top with the $SU(2)$ self-rotating 
in the isospin direction. The Yang's $SU(2)$ monopole harmonics 
\cite{Yang} can be interpret as the wave functions in the coherent state 
picture in the physical gauge. 

Based on the orthogonality and completeness of the state vectors $\left | 
\begin{array} {l}
R \\
J \end{array} \right \rangle$ belonging to an irreducible representation 
$R = [r_1, r_2 = I]$ of $SO(5)$, we can give the completeness condition of 
the coherent state vectors
\begin{equation}
\frac{D(r_1, r_2 = I)}{A(\Omega)} \int d \Omega | \Omega , R 
\rangle \langle \Omega , R| = 1,
\end{equation}
where $A(\Omega) = Area(S^{4} \times S^{3}) = \frac{1}{12} (8 \pi)^2)^2$. 
Every irreducible representation of $SO(5)$ corresponds to a complete set 
of the coherent state vectors. The coherent states of the different 
irreducible representations are orthogonal each other. The coherent state 
corresponding to the LLL states is very important for the description of 
non-commutative geometry of 4-dimensional quantum Hall droplet. 

In order to avoid the label of the irreducible representation of the LLL 
states confusing with the parameter $I$ of the model, we denote the 
irreducible representation $R=[r_1 = I,r_2 = I]$ as $\frac{P}{2} \equiv 
[r_1 = I,r_ 2 = I]$. The LLL degeneracy states consist of the Hilbert 
space ${\cal H}^{(\frac{P}{2})}$. Because of the LLL states are the lowest 
energy states of particle's living in, for a given $I$, the smallest 
admissible irreducible representation of $SO(5)$ is $R = [r_1 = I, r_2 = 
I] = \frac{P}{2}$. Therefore, the irreducible representations of $SO(5)$ 
are truncated since there exists an Yang's $SU(2)$ monopole. If we focus 
on the Hilbert space of the LLL states, we can determine the matrix forms 
of tensor operator for the $SU(2)$ monopole harmonics, which are the 
$D(r_1 = I, r_2 = I) \times D(r_1 = I,r_2 = I)$ matrices. The coupling 
relation between the tensor operators provides a truncated parameter for 
the tensor operator for the $SU(2)$ monopole harmonics. The number of 
independent operators is $(D(r_1 = I, r_2 = I))^2$, which we shall explain 
in more detail in the next section. Therefore, we should replace the 
functions by the $D(r_1 = I,r_2 = I) \times  D(r_1 = I, r_2 = I)$ matrices 
on the fuzzy $S^{4}$. Exactly, the cross sections in the fibre bundle 
$SO(5)/SU(2)$, which is a bundle over $S^{4}$ with fibre $S^{3}$, are 
replaced by the $D(r_1 = I,r_2 = I) \times  D(r_1 = I,r_2 = I)$ matrices 
on the fuzzy $S^{4}$. Thus, the algebra on the fuzzy $S^{4}$ becomes 
non-commutative. The direct product of $N$ single-particle Hilbert spaces 
${\cal H}^{(\frac{P}{2})}$ can be used to build up the Hilbert space of 
4-dimensional quantum Hall droplet. Hence, the fuzzy $S^{4}$ appears 
naturally in the description of particle's moving on the $S^{4}$ with the 
Yang's $SU(2)$ monopole at the center of the sphere. Consequently, the 
fuzzy $S^{4}$ is the description of non-commutative geometry of 
4-dimensional quantum Hall droplet.

\section{Fuzzy monopole harmonics and Matrix operators of fuzzy $S^{4}$}

The construction of fuzzy $S^{4}$ is to replace the functions on $S^{4}$ 
by the non-commutative algebra taken in the irreducible representations of 
$SO(5)$. This is a full matrix algebra which is generated by the fuzzy 
monopole harmonics. These fuzzy monopole harmonics consist of a complete 
basis of the matrix space. In this section, we shall find the explicit 
forms of such fuzzy monopole harmonics by means of the expressions of the 
$SU(2)$ monopole harmonics given in the previous section. Our main task of 
this section is to construct the operators corresponding to the $SU(2)$ 
monopole harmonics, and to give the matrix elements of these operators 
acting on the LLL states.

Although for a given $I$, the $SU(2)$ monopole harmonics of the smallest 
admissible irreducible representations of $SO(5)$ can be regarded as the 
single-particle wave functions of the LLL in 4-dimensional quantum Hall 
system, the $SU(2)$ monopole harmonics smaller than the smallest admissible 
irreducible representations of $SO(5)$ are useful for us to construct the 
fuzzy monopole harmonics on $SO(5)/SU(2)$. Such $SU(2)$ monopole harmonics 
can be read off from the expressions obtained in the previous section by 
the changing of the parameter $I$. If we replace $I$ by $J$, the $SO(5)$ 
irreducible representation of the monopole harmonics becomes $R = [r_1, 
r_2 = J]$. Equivalently, we can use $R = [r_1, r_2]$ and $J^{(0)} \equiv 
\left ( \begin{array} {ll}
0 & r_2 \\
0 & r_{2 z} \end{array} \right )$ to label them. Thus, the $SU(2)$ 
monopole harmonics are generally expressed as ${\cal D}_{J 
J^{(0)}}^{(R)}(\Omega)$, which can be obtained by replacing $I^{(0)}$ with 
$J^{(0)}$ in the equation (8). We can find their corresponding coherent 
states by the same replacement. 

By using the standard techniques of the generalized coherent state 
\cite{Peremolov}, we can now construct the operator ${\hat {\cal 
Y}}^R_J$ corresponding to the $SU(2)$ monopole harmonics ${\cal D}_{J 
J^{(0)}}^{(R)}(\Omega)$. Noticing that this operator is an operator of 
acting in the LLL Hilbert space and ${\cal D}_{J J^{(0)}}^{(R)}(\Omega)$ 
is regarded as the basic function, we can express it as the following form
\begin{equation}
{\hat {\cal Y}}^R_J = \frac{D(r_1 = I, r_2 = I)}{A(\Omega)} \int d \Omega 
\sum_{r_{2 z}} \langle r_2, r_{2 z} |{\cal D}_{J J^{(0)}}^{(R)}(\Omega) | 
\Omega , \frac{P}{2} \rangle \langle \Omega , \frac{P}{2} |.
\end{equation}
The matrix elements of this operator in the LLL Hilbert space are read as
\begin{eqnarray}
\left \langle \begin{array} {l}
\frac{P}{2} \\ 
K^1 \end{array} \right | {\hat {\cal Y}}^{R}_{J}
\left | \begin{array} {l}
\frac{P}{2} \\
K^2 \end{array} \right \rangle & = & 
\frac{D(r_1 = I, r_2 = I)}{A(\Omega)}\nonumber\\
& \times & \int d \Omega \sum_{I_{1 z}, r_{2 z}, I_{2 z}} {\cal D}_{K^1 
I^{(0)}}^{*(\frac{P}{2})}(\Omega) {\cal D}_{J J^{(0)}}^{(R)}(\Omega) {\cal 
D}_{K^2 I^{(0)}}^{(\frac{P}{2})}(\Omega) \langle I, I_{1 z} | r_2, r_{2 z} 
\rangle | I, I_{2 z} \rangle.
\end{eqnarray}

The above integral can be analytically performed by making use of the 
integral formulae about the Jacobi polynomial and the product of three 
$SU(2)$ D-functions, and the properties of $SU(2)$ D-function. The result 
is
\begin{eqnarray}
\left \langle \begin{array} {l}
\frac{P}{2} \\
K^1 \end{array} \right | {\hat {\cal Y}}^R_J \left | \begin{array} {l}  
\frac{P}{2} \\
K^2 \end{array} \right \rangle & = & \left \langle \begin{array} {l}  
\frac{P}{2} \\
K^1 \end{array} \right \| \begin{array} {l}
R \\
J \end{array} \left \| \begin{array} {l}
\frac{P}{2} \\
K^2 \end{array} \right \rangle \left \{ \begin{array} {lll}
k^1_1 & j_1 & k^2_1 \\  
I  & r_2 & I \\  
k^1_2 & j_2 & k^2_2 \end{array} \right \} \nonumber\\
& \times & \langle k^1_1, k^1_{1 z} | j_1, 
j_{1 z}; k^2_1, k^2_{1 z} \rangle \langle k^1_2, k^1_{2 z} | j_2, j_{2 z}; 
k^2_2, k^2_{2 z} \rangle,
\end{eqnarray}
where $\left \langle \begin{array} {l}
\frac{P}{2} \\
K^1 \end{array} \right \| \begin{array} {l}
R \\
J \end{array} \left \| \begin{array} {l}
\frac{P}{2} \\
K^2 \end{array} \right \rangle$ is independent of the $SU(2)$ magnetic 
quantum numbers, and is composed of two parts, i.e., it is equal to
$\Phi(k_1^1, k_2^1, k_1^2, k_2^2, j_1, j_2, I, r_2) \Theta(k_1^1, k_2^1, 
k_1^2, k_2^2, j_1, j_2, I, r_1, r_2)$. The part of $\Phi$ is from the 
contribution of the integral with respect to the variables $\alpha, 
\beta, \gamma, \alpha_I, \beta_I $ and $\gamma_I $, which is provided by 
the normalized coefficients and the arisen factors when we expressed the sum over all 
$SU(2)$ magnetic quantum numbers of six $SU(2)$ coupling coefficients as 
the 9-j symbol. Its expression is given by
\begin{equation}
\Phi = (-1)^{ j_1 + j_2 + j_1 + k_1^1 + k_1^2 }
\sqrt{\frac{(2 I + 1)^3 (2 k_2^1 + 1)^2 (2 k_2^2 + 1)(2 k_1^2 + 1)(2r_2 +1)(2 j_1 + 1)(2 j_2 + 1)}{(8 \pi^2 )^2}} .
\end{equation} 
The another part $\Theta$ is given by the integrated part of 
$\theta$. It is read as
\begin{eqnarray}
\Theta & = &\sqrt{\frac{(2I + 1)^2 (2 r_1 + 1)}{8}} \int_{0}^{\pi} d\theta  
d_{k_1^1 -k_2^1 , -I-1}^{(I+1)} (\theta ) d_{k_1^2 -k_2^2 , -I-1}^{(I+1)} (\theta )
d_{j_1 -j_2 , -j_1 -j_2 -1}^{(r_1 +1)} (\theta ) \nonumber\\ 
& = & \{2^{2(k_1^1 + k_1^2 + j_1) - 2(k_2^1 + k_2^2 + j_2) + 
3}\}^{-\frac{1}{2}} \nonumber\\
& \times &\{ \frac{(I + k_1^1 - k_2^1 + 1)! (I + k_1^2 - k_2^2 + 1)! (I + 
k_2^1 - k_1^1 + 1)! (I + k_2^2 - k_1^2 + 1)!} {(2I+3)^2 (2 r_1 + 3)((2 I + 
2)!)^2} \}^{-\frac{1}{2}} \nonumber \\ 
& \times & \{\frac{(r_1 + j_1 + j_2 + 2)! (r_1 - j_1 - j_2)!}{(r_1 + 1 + 
j_2 - j_1)! (r_1 + 1-j_2 + j_1)!} \}^{-\frac{1}{2}} \frac{D(r_1 = I, r_2 = 
I)}{A(\Omega)} \nonumber \\
& \times & \frac{2^{2 I + j_1 - j_2 -2 k_2^1 - 2 k_2^2} \Gamma(j_1 + 
k_1^1 + k_1^2 + 2) \Gamma(k_2^1 + k_2^2 - j_2 + 1) \Gamma(r_1 + j_1 + j_2 
+ 3)}{(r_1 - j_1 + j_2 + 1)! \Gamma(2 I + 3 + j_1 - j_2)\Gamma(2 j_1 + 2)} 
\nonumber \\
& \times & ~_3F_2(- r_1 + j_1 - j_2 - 1, r_1 + j_1 - j_2 + 2, j_1 + k_1^1 
+ k_1^2 + 2; 2 j_1 + 2, 2 I + 3 + j_1 - j_2; 1),
\end{eqnarray}
where $~_{3}F_{2}$ is the hyper-geometric function. 

Although the 9-j symbol of $SO(3)$ is also independent of the $SU(2)$ 
magnetic quantum numbers, $\Phi$, $\Theta$ and it are very important for 
the matrix form of the operator since they all depend on the $SO(4)$ 
subgroup quantum numbers of $SO(5)$. Hence, they may not be smeared out by 
re-scaling. In particular, the 9-j symbol together with two $SU(2)$ 
coupling coefficients will provide the selection rules of the matrix 
elements of the operator ${\hat {\cal Y}}^R_J$. From the coupling 
relations of 9-j symbol's elements and the triangle relations of $SU(2)$ 
angular momentum coupling, we find that $r_2$ can be evaluated in the 
range zero to $2 I$. Furthermore, because of $k_1^1 + k_2^1 = k_1^2 + 
k_2^2 = I$, the highest values of these $k$'s is $I$. We know that the 
maximums of $j_1$ and $j_2$ both are $2 I$ from the coupling relations of 
two $SU(2)$ coupling coefficients. The scheme of the $SO(5)$ irreducible 
representation containing the $SO(4)$ blocks \cite{Yang} tells us that the 
largest admissible value of $r_1$ is $2 I$. Acting in the LLL Hilbert 
space, the operators ${\hat {\cal Y}}^R_J$ producing the nonzero 
contributions are those belonging to the irreducible representations of 
$SO(5)$ of $0 \leq r_2 \leq r_1 \leq 2 I$.

The number of such operators can be calculated by means of the dimension 
formula of irreducible representation of $SO(5)$. From the equation (2), 
we have
\begin{eqnarray}
\sum_{0 \leq r_2 \leq r_1 \leq 2 I} D(r_1, r_2) & = & \frac{1}{6}\sum_{0 
\leq r_2 \leq r_1 \leq 2 I} (1 + r_1 - r_2)(1 + 2 r_2)(2 + r_1 + r_2)(3 + 
2 r_1) \nonumber \\
& = &\frac{1}{36} (2 I + 1)^2 (2 I + 2)^2 (2 I + 3)^2.
\end{eqnarray}
This is exactly equal to the square of the dimensionality of the LLL 
Hilbert space. This show that the operators
\begin{equation}
{\hat {\cal Y}}^R_J,~~~~~~~~~~~~~~~~~~0 \leq r_2 \leq r_1 \leq 2 I
\end{equation}
constitute a complete set of irreducible representations of $SO(5)$ in the 
matrix algebra with the square of the dimension of the LLL Hilbert space. 
We also call them as the fuzzy monopole harmonics.

Subsequently, we shall discuss the coupling of the $SO(5)$ angular 
momentums. It is helpful to explore the Hilbert space structure of 
many-body system corresponding to the system (1), in especial, of the 
4-dimensional quantum Hall system. The Kronecker product of two unitary 
irreducible representations of a semi-simple group is completely 
reducible, but it is generally not simply reducible. Hence, a given 
representation may appear more than once in the decomposition of the 
Kronecker product. However, in the case of $SO(5)$ angular momentum, every admissible 
irreducible representation only appears once in the decomposition of the 
Kronecker product \cite{Yang}. This conclusion is a  corollary of the theorems $2$, 
$4$ and $8$ in the reference \cite{Yang}. Based on this fact, we have the following 
coupling relation of the state vectors
\begin{equation}
\left | \begin{array} {l}
R \\
J \end{array} \right \rangle = \sum_{J_1, J_2} \left \langle 
\begin{array}{ll}
R_1 & R_2 \\  
J_1 & J_2 \end{array} \right. \left | \begin{array}{l}
R \\
J \end{array} \right \rangle \left | \begin{array}{ll}
R_1 & R_2 \\  
J_1 & J_2 \end{array} \right \rangle
\end{equation}
Then, for the case of $SO(5)$ angular momentum, the decomposition of the Kronecker product 
of two $SO(5)$ irreducible representations is accomplished by the $SO(5)$ coupling coefficients 
$\left \langle \begin{array}{ll}
R_1 & R_2 \\
J_1 & J_2 \end{array} \right. \left | \begin{array}{l}
R \\
J \end{array} \right \rangle$. These coupling coefficients obey 
the usual 
orthogonality relations. We again emphasize that here all state vectors 
$\left | \begin{array} {l}
R \\
J \end{array} \right \rangle $ belong to the Hilbert space ${\cal H}^{(R)}$, which
is composed of the monopole harmonics ${\cal D}_{J J^{(0)}}^{(R)}(\Omega)$. 

In the previous section, we have explained that $\sum_{r_{2z}} {\cal D}_{J 
J^{(0)}}^{*(R)}(\Omega) | r_{2}, r_{2z} \rangle $ can be regarded as the wave 
function with the $U(1)$ gauge degree of freedom generated by the finite 
rotation of $SO(5)$ acting on  the fixed state vector in the Hilbert 
space, of course, $\sum_{r_{2z}} \langle r_{2}, r_{2z} |{\cal D}_{J 
J^{(0)}}^{(R)}(\Omega)$ also can be done. By using this property of the wave 
functions and the orthogonality relations of the coupling coefficients, we find that
\begin{eqnarray}
\sum_{R, J, r_{2 z}, r_{2 z}^1, r_{2 z}^2} \langle r_2, r_{2 z} |{\cal D}_{J 
J^{(0)}}^{(R)}(\Omega) \left \langle 
\begin{array}{l}
R \\
J^{(0)} \end{array} \right. \left | \begin{array}{ll}
R_1 & R_2 \\  
J_1^{(0)} & J_2^{(0)} \end{array} \right \rangle \left \langle 
\begin{array}{ll}
R_1 & R_2 \\
J_1 & J_2 \end{array} \right. \left | \begin{array}{l}
R \\
J \end{array} \right \rangle \nonumber\\
= \sum_{r_{2 z}^1} \langle r^1_2, r^1_{2 z} |{\cal D}_{J_1 
J_1^{(0)}}^{(R_1)}(\Omega) \sum_{r_{2 z}^2} \langle r^2_2, r^2_{2 z} | 
{\cal D}_{J_2 J_2^{(0)}}^{(R_2)}(\Omega) ,
\end{eqnarray}
where $R_i = [r_1^i, r_2^i]$ , and $J_i^{(0)} = \left ( \begin{array}{ll}  
0 & r_2^i \\
0 & r_{2 z}^i \end{array} \right )$ for $i=1, 2, 3$ (see below). By means of 
the orthogonality and normalized condition of the $SU(2)$ monopole 
harmonic, we can obtain the explicit expression of the coupling 
coefficients of the $SO(5)$ angular momentum 
\begin{eqnarray}
\left \langle \begin{array}{ll}
R_2 & R_3 \\
J_2 & J_3 \end{array} \right. \left | \begin{array}{l}
R_1 \\
J_1 \end{array} \right \rangle 
& = & 
{\langle R_1 \| R_2 \| R_3 \rangle }^{-1} \nonumber\\
& \times & \int d \Omega \sum_{r_{2z}^1 , r_{2 z}^2 , r_{2 z}^3 } {\cal D}_{J_1 
J_{1}^{(0)}}^{*(R_1 )}(\Omega) {\cal D}_{J_2 J_2^{(0)}}^{(R_2 )}(\Omega) {\cal 
D}_{J_3 J_3^{(0)}}^{(R_3 )}(\Omega) \langle r_2^1 , r_{2z}^1 | r_2^2 , r_{2z}^2 
\rangle | r_2^3 , r_{2z}^3 \rangle ,
\end{eqnarray}
where $\langle R_1 \| R_2 \| R_3 \rangle = \sum_{r_{2 z}^1, r_{2 z}^2, 
r_{2 z}^3} \left \langle \begin{array}{l}
R_1 \\
J_1^{(0)} \end{array} \right. \left | \begin{array}{ll}
R_2 & R_3 \\  
J_2^{(0)} & J_3^{(0)} \end{array} \right \rangle$ is invariant under the 
rotation transformation of $SO(5)$, hence is a pure scale factor. 

Performing the calculation of the above integral, we get
\begin{eqnarray}
\left \langle \begin{array}{ll}
R_2 & R_3 \\
J_2 & J_3 \end{array} \right. \left | \begin{array}{l}
R_1 \\
J_1 \end{array} \right \rangle & = & \frac{\left \langle \begin{array}{l}
R_1 \\
J_1 \end{array} \right \| \begin{array}{l}
R_2 \\
J_2 \end{array} \left \| \begin{array}{l}
R_3 \\
J_3 \end{array} \right \rangle}{\langle R_1 \| R_2 \| R_3 \rangle} \left 
\{ \begin{array}{lll}
j^1_1 & j^2_1 & j^3_1 \\
r^1_2 & r^2_2 & r^3_2 \\
j^1_2 & j^2_2 & j^3_2 \end{array} \right \} \nonumber \\
& \times & \langle j^1_1, j^1_{1 z} | j^2_1, j^2_{1 z}; j^3_1, j^3_{1 z} 
\rangle \langle j^1_2, j^1_{2 z} | j^2_2, j^2_{2 z}; j^3_2, j^3_{2 z} 
\rangle,
\end{eqnarray}
Similar to the matrix elements of the fuzzy monopole harmonics, $\left \langle 
\begin{array}{l}
R_1 \\
J_1 \end{array} \right \| \begin{array}{l}
R_2 \\
J_2 \end{array} \left \| \begin{array}{l}
R_3 \\
J_3 \end{array} \right \rangle = \tilde{\Theta}\tilde{\Phi}$. $\tilde{\Theta}$ and
$\tilde{\Phi}$ are simply similar to $\Theta$ and $\Phi$ respectively. In 
$\tilde{\Phi}$, it is only different 
of the replacing the quantum numbers in $\Phi$ with the corresponding quantum 
number at present. The expression of $\tilde{\Phi}$ is given by
\begin{equation}
\tilde{\Phi}= (-1)^{j_1^2 -j_1^3 +j_2^2 -j_2^3 +r_2^2 -r_2^3 }
\sqrt{\frac{(2r_2^1 + 1)(2 j_2^1 + 1)}{(8 \pi^2 )^2 (2j_1^1 +1)}}
\prod_{i = 1}^{3} (-1)^{j_1^i -r_2^i} \sqrt{(2r_2^i +1)(2j_1^i +1)(2j_2^i +1)}.
\end{equation} 
The expression of $\tilde{\Theta}$ is given by the following integration
\begin{eqnarray}
\tilde{\Theta} & = & \int_{-1}^{1} d x (1 - x^2)^{-\frac{1}{2}} \prod_{i = 
1}^{3} [\frac{2^{2 j_1^i - 2 j_2^i + 1}(r_1^i + j_1^i + j_2^i + 2)! (r_1^i 
- j_1^i - j_2^i)!}{(2r_1^i+3)(r_1^i - j_1^i + j_2^i + 1)! (r_1^i + j_1^i - 
j_2^i + 1)!}]^{- \frac{1}{2}} \nonumber\\
& \times & (1 - x)^{\frac{2 j_1^i + 1}{2}} (1 + x)^{- \frac{2 j_2^i + 
1}{2}} P^{(2 j_1^i + 1, - 2 j_2^i - 1)}_{r_1 - j_1^i + j_2^i + 1}(x).
\end{eqnarray}

Performing the analysis parallel to the selection rules of the matrix 
elements, we find that the coupling coefficient vanishes unless the 
$r_1^i$ and $r_2^i$ satisfy the  generalized triangular conditions as 
following
\begin{equation}
|r_2^1 - r_2^2| \leq r_2^3 \leq r_1^3 \leq r_1^1 + r_1^2, 
\end{equation}
etc. The selection rules of the $SO(5)$ coupling coefficients about the 
$SO(4)$ subgroup quantum numbers are provided by the 9-j symbol and two 
$SU(2)$ coupling coefficients.

Usually, the evaluation of a coupling coefficient involving a set of basis 
states labeled by the irreducible representations of a chain of nested 
subgroup for a semi-simple group. In this approach, the most important 
ingredient is Racah's  factorization lemma, which ensure the coupling 
coefficient to factorize into the coupling coefficients of the subgroup. 
Then, the evaluation of a coupling coefficient becomes the calculations of 
the re-coupling coefficients, also called the isoscalar factors. In fact, 
the calculation of the  isoscalar factor is very difficult. The advantage 
of this method is that it can overcome the trouble with a semi-simple 
group being not generally simply reducible. However, for the case of $SO(5)$
angular momentum, the group $SO(5)$ is simply reducible in the $SO(5)$ level, 
i.e., every irreducible representation of $SO(5)$ only appears once in the 
decomposition of the direct product of two $SO(5)$ irreducible representations. 
Our deriving the explicit expression of the coupling coefficients of $SO(5)$ is just 
based on this fact. Since the expression of the coupling coefficients 
given here is analytically exact, it is not necessary further to 
calculate the isoscalar factors appearing in the factorization of the 
coupling coefficients of the chain of nested subgroup. 

In fact, $\tilde{\Theta}$ and $\tilde{\Phi}$ 
together with the 9-j symbol of $SO(3)$ in the Eq.(23) provide the isoscalar factor
for $SO(5)\supset SU(2)\times SU(2)$ in the level of $SO(5)$ angular momentum. Here, 
$SU(2)\times SU(2)$ is the relatively direct product, of which the rotation transformation 
is generated by one part ${\hat J}_{1i}$ and another relative part ${\hat J}_{2i} =
{\hat J}_{2i}^{(0)} +{\hat I}_{i}$ by means of $\sum_i  {\hat J}_{1i}^2 =\sum_i  
{\hat J}_{2i}^{(0)2}$. Furthermore, we give here the analytically exact expressions 
of all isoscalar factors of  $SO(5)$ in the $SO(5)$ angular momentum level. It is emphasized 
worthily that all irreducible representations of $SO(5)$ discussed by us here are those 
belonging to the $SO(5)$ angular momentum, exactly, those corresponding to the Hilbert
spaces which are composed of the $SU(2)$ monopole harmonics. 
   
We can construct many-particle's states from the single-particle states by 
using of the expression of the coupling coefficients. In the procedure of 
many-particle state construction, the re-coupling coefficients can be 
obtained by performing summation over products of the coupling 
coefficients. Therefore, the expression of the coupling coefficients given 
by us here is important for the study of the 4-dimensional quantum Hall 
system and the physical system with the rotation symmetry of $SO(5)$. In 
fact, the calculation of the matrix elements of fuzzy monopole harmonics 
is that of the coupling coefficients.

On the other hand,  the matrix operators ${\hat {\cal Y}}^R_J, 0 \leq r_2 \leq r_1 \leq 2 I$
are the single-body operators acting in the LLL Hilbert space since the matrix forms 
of them are provided by the matrix elements between the LLL states of single particle system. 
By determining the ground states, i.e., the LLL states, which can be obtained by the linear 
combinations of these operators acting on the vacuum state, we can use these operators to generate 
all possible single particle states including the quasi-particle and quasi-hole states near the LLL 
states. In this sence, the fuzzy monopole harmonics given by us are the generatores of the general 
wave functions of single particle system. In order to explore the detail structure of this  system's 
Hilbert space and the properties of quantum states and operators in the 4-diemnsional quantum Hall 
droplet, we need further to discuss the algebrical structure of these operators, i.e., the commutation 
relations between them.

\section{Matrix algebra of fuzzy $S^{4}$ and non-commutative geometry from 
the system}

\indent

Now, we turn to the discussion of the matrix algebra of the fuzzy monopole 
harmonics. It is clear that the matrix operators ${\hat {\cal Y}}^R_J, 0 
\leq r_2 \leq r_1 \leq 2 I$ belong to the irreducible representations of 
$SO(5)$. All operators belong to the irreducible representations of 
$SO(5)$ must satisfy the following relation of the tensor products of the 
operators
\begin{equation}
{\hat {\cal T}}_{J_1}^{R_1} {\hat {\cal T}}_{J_2}^{R_2} = \sum_{R, J} 
\left \langle \begin{array}{l}
R \\
J \end{array} \right. \left | \begin{array}{ll}
R_1 & R_2 \\
J_1 & J_2 \end{array} \right \rangle {\hat {\cal T}}_J^R.
\end{equation}

Another useful relation is the Wigner-Eckart theorem for the theory of 
$SO(5)$ angular momentum. It is read as
\begin{equation}
\left \langle \begin{array}{l}
R_1 \\
J_1 \end{array} \right | {\hat {\cal T}}_J^R \left | \begin{array}{l}
R_2 \\
J_2 \end{array} \right \rangle = \left \langle \begin{array}{ll}
R_1 & R \\
J_1 & J \end{array} \right. \left | \begin{array}{l}
R_2 \\
J_2 \end{array} \right \rangle \langle R_1 \| {\hat {\cal T}}^R \| R_2 
\rangle,
\end{equation}
where $\langle R_1 \| {\hat {\cal T}}^R \| R_2 \rangle$ is independent of 
the subgroup $SO(4)$ quantum numbers of $SO(5)$, which is called as the 
reduced matrix elements of the operator ${\hat {\cal T}}_J^R$. This 
theorem can be derived by using of the standard method of the group 
representation \cite{Wybourne} and the fact that every irreducible 
representation of $SO(5)$ appears once in the decomposition of the 
Kronecker product of two irreducible representations of $SO(5)$ in 
the case of $SO(5)$ angular momentum. 

We can obtain the relation between the reduced matrix elements by 
calculating the matrix elements of the tensor product's relation of the 
operators. Furthermore, we scale the operators ${\hat {\cal Y}}_J^R$ into
\begin{equation}
{\tilde {\hat {\cal Y}}}_J^R = \frac{{\hat {\cal Y}}_J^R}{\langle 
\frac{P}{2} \| {\hat {\cal Y}}^R \| \frac{P}{2} \rangle}.
\end{equation}
By means of the relation between the reduced matrix elements, we can now 
read off the matrix algebraic relation of the operators ${\tilde {\hat 
{\cal Y}}}_J^R$ from the operator product relation (27). This 
relation is
\begin{equation}
{\tilde {\hat {\cal Y}}}_{J_1}^{R_1} {\tilde {\hat {\cal Y}}}_{J_2}^{R_2} 
= \sum_{R, J} \left \langle \begin{array}{l}
R \\
J \end{array} \right. \left | \begin{array}{ll}
R_1 & R_2 \\
J_1 & J_2 \end{array} \right \rangle \left \{ \begin{array}{lll}
R & R_1 & R_2 \\
\frac{P}{2} & \frac{P}{2} & \frac{P}{2} \end{array} \right \}
[D(\frac{P}{2})]^{-1} {\tilde {\hat{\cal Y}}}_J^R,
\end{equation}
where we have define that $D(R)=D(r_1, r_2)$ and $D(\frac{P}{2}) =D(r_1 = 
I, r_2 = I)$. The operator of the l.h.s. of the Eq.(30) is given by the matrix 
multiplication between the operators ${\tilde {\hat {\cal Y}}}_{J_1}^{R_1}$ and 
${\tilde {\hat {\cal Y}}}_{J_2}^{R_2}$, which act in the LLL Hilbert space. The 
6-J symbol of $SO(5)$ is given by
\begin{eqnarray}
\left \{ \begin{array}{lll}
R & R_1 & R_2 \\
\frac{P}{2} & \frac{P}{2} & \frac{P}{2} \end{array} \right \} & = & 
\sum_{J, J_1, J_2, K, K_1, K_2} \left \langle \begin{array}{ll}
R_1 & R_2 \\
J_1 & J_2 \end{array} \right. \left | \begin{array}{l}
R \\
J \end{array} \right \rangle \left \langle \begin{array}{l}
\frac{P}{2} \\
K_2 \end{array} \right. \left | \begin{array}{ll}
R & \frac{P}{2} \\
J & K_1 \end{array} \right \rangle \nonumber\\
& & \times \left \langle \begin{array}{ll}
\frac{P}{2} & R_1 \\
K_1 & J_1 \end{array} \right. \left | \begin{array}{l}
\frac{P}{2} \\
K \end{array} \right \rangle \left \langle \begin{array}{ll}
\frac{P}{2} & R_2 \\
K & J_2 \end{array} \right. \left | \begin{array}{l}
\frac{P}{2} \\
K_2 \end{array} \right \rangle.
\end{eqnarray}
It can be seen easily that the 6-J symbol of $SO(5)$ is possessed of the 
properties very similar to the 6-j symbol of $SO(3)$. 

The matrix algebra (30) of the fuzzy $S^{4}$  is formally analogous to 
that of the fuzzy $S^{2}$ \cite{Alekseev,Chan,Hou}. Because of the 
operators ${\tilde {\hat{\cal Y}}}_J^R, 0 \leq r_2 \leq r_1 \leq 2 I$ 
consist of a complete basis of the $D(\frac{P}{2}) \times  D(\frac{P}{2})$ 
matrix algebra, any operator acting in the LLL Hilbert space can be 
expressed as a linear combinationof the matrices ${\tilde {\hat {\cal 
Y}}}_J^R, 0 \leq r_2 \leq r_1 \leq 2 I$. The  product of such operators 
again becomes the linear combination of the matrices due to  the matrix 
algebraic relation (30). In the other words, the matrix algebra is a 
fundamental relation to determine the algebraic relations of all operators 
acting in the LLL Hilbert space. The origin of the matrix operators and 
their algebra appearing in the 4-dimensional quantum Hall system is due to 
the existence of the Yang's $SU(2)$ monopole in the system. Its appearance 
makes the coordinate space $S^{4}$, which the particles live in, become 
non-commutative. The description of this non-commutative geometry can be 
obtained by replacing the algebra of the functions with the matrix 
operators, i.e., the fuzzy monopole harmonics. The procedure of our 
finding these matrix operators and their matrix algebra above is just 
establishing the description of this non-commutative geometry. In this 
sense, the matrix algebra given here is the algebra of fuzzy $S^{4}$, and 
describes the non-commutativity of the coordinate space $S^{4}$. When the 
strength of the Yang's $SU(2)$ monopole vanishes, i.e., $I=0$, the 
dimension of the matrix becomes one dimensional and trivial, and then the 
matrix algebra does the algebra of the functions. Consequently, the fuzzy 
$S^{4}$ becomes a classical $S^{4}$ when $I = 0$. 

The matrix algebra given by us is an associative algebra. The simple 
interpretation of the associativity of the matrix algebra is that since 
these matrices are the operators acting in the LLL Hilbert space, they can 
be expressed as
\begin{equation}
{\tilde {\hat{\cal Y}}}^R_J = \sum_{K^1, K^2} \left \langle 
\begin{array}{l}
\frac{P}{2} \\
K^1 \end{array} \right | {\tilde {\hat{\cal Y}}}^R_J \left | 
\begin{array}{l}
\frac{P}{2} \\
K^2 \end{array} \right \rangle \left | \begin{array}{l}
\frac{P}{2} \\
K^1 \end{array} \right \rangle \left \langle \begin{array}{l}
\frac{P}{2} \\
K^2 \end{array} \right |.
\end{equation}
Thus, the products of three operators become those of three matrices. The 
associativity of the matrix algebra is equivalently described by the 
identity of the matrices
\begin{eqnarray}
\sum_{K^\prime, K^{\prime \prime}} \left \langle \begin{array}{l}  
\frac{P}{2} \\
K^1 \end{array} \right | {\tilde {\hat{\cal Y}}}^{R_1}_{J_1} \left | 
\begin{array}{l}
\frac{P}{2} \\
K^\prime \end{array} \right \rangle \left \langle \begin{array}{l}
\frac{P}{2} \\
K^\prime \end{array} \right | {\tilde {\hat {\cal Y}}}^{R_2}_{J_2} 
\left | \begin{array}{l}
\frac{P}{2} \\
K^{\prime \prime} \end{array} \right \rangle \left \langle 
\begin{array}{l}
\frac{P}{2} \\
K^{\prime \prime} \end{array} \right | {\tilde {\hat {\cal 
Y}}}^{R_3}_{J_3} \left | \begin{array}{l}
\frac{P}{2} \\
K^2 \end{array} \right \rangle \nonumber \\
= \sum_{K^\prime, K^{\prime \prime}} \left \langle \begin{array}{l} 
\frac{P}{2} \\
K^1 \end{array} \right | {\tilde {\hat{\cal Y}}}^{R_1}_{J_1} \left | 
\begin{array}{l}
\frac{P}{2} \\
K^{\prime \prime} \end{array} \right \rangle \left \langle 
\begin{array}{l}
\frac{P}{2} \\
K^{\prime \prime} \end{array} \right | {\tilde {\hat {\cal 
Y}}}^{R_2}_{J_2} \left | \begin{array}{l}
\frac{P}{2} \\
K^\prime \end{array} \right \rangle \left \langle \begin{array}{l}
\frac{P}{2} \\
K^\prime \end{array} \right | {\tilde {\hat {\cal Y}}}^{R_3}_{J_3} \left | 
\begin{array}{l}
\frac{P}{2} \\
K^2 \end{array} \right \rangle.
\end{eqnarray}
Obviously, the results of the above summations are same because there does 
not exist any singularity to make the summarizing sequences change the 
results of the summations. Of course, one can straightforwardly show the 
associativity of the operator algebra of  the fuzzy $S^{4}$ in the similar 
manner of the proof of the associativity for the case of fuzzy $S^{2}$ 
\cite{Alekseev}. This is a more complicate and more technical procedure in 
which the generalized Biedenharn-Elliott relation for the 6-J symbols of 
$SO(5)$ should be established. Although this generalized relation is very 
useful, we do not discuss it here.

From the above discussions, one can see that the matrix algebra (30) is the 
multiplicative relation of the matrices produced by the fuzzy monopole harmonics 
acting in the LLL Hilbert space of single particle system. These matrices consist 
of a complete set of all operators belonging to the LLL Hilbert space of
single particle system. It is well known that the Hilbert space of the many 
particles can be constructed from  the single particle's Hilbert spaces by the 
coupling of some few of single-particles. The most simplest way of fuzzy monopole 
harmonics'construction of the system of $N$ particles is to construct first the most 
elementary fuzzy monopole harmonics of the system of $N$ particles by making of the 
symmetrical direct sum of $N$ single particle fuzzy monopole harmonics. Then, one can produce 
all fuzzy monopole harmonics of  the system of $N$ particles by using of the matrix algebra (30). 
This way can not provide the generally truncated rule of  the irreducible representations of 
$SO(5)$ corresponding to the fuzzy monopole harmonics, which are composed of the LLL Hilbert 
space of  the system of $N$ particles. On the other hand, it is not obvious that the connection 
between the construction of the fuzzy monopole harmonics in this way and the wave functions of 
quasi-particle or quasi-hole excitations in the Laughlin's and Haldane's forms. We shall give another 
scheme of the construction of fuzzy monopole harmonics of the system of $N$ particles in the rest of 
this section, and the most simplest way is a special case of the following scheme. 

Subsequently, we shall discuss how the fuzzy 
monopole harmonics of the coupling system are constructed from the 
subsystems composed of the coupling system. Since the particles spread 
over the four-sphere $S^{4}$, this study is very important for the 
description of non-commutative geometry of the 4-dimensional quantum Hall 
droplet. Our starting point is the tensor product relation of two tensor 
operators belonging to the irreducible representations of $SO(5)$ which correspond 
to the Yang's $SU(2)$ monopole harmonics. The 
tensor operators ${\hat {\cal T}}^{R^1}_{J^1}(1)$ and ${\hat {\cal 
T}}^{R^2}_{J^2}(2)$ are supposed to work on the subsystem $1$ and the 
subsystem $2$ respectively of a system in order to distinguish them 
labelled by adding to $1,2$. Because of these operators belonging to the 
irreducible representations of $SO(5)$, they must satisfy the tensor 
product relation of the operators
\begin{equation}
{\hat {\cal T}}_J^R = \sum_{J^1, J^2} \left \langle \begin{array}{ll}
R^1 & R^2 \\
J^1 & J^2 \end{array} \right. \left | \begin{array}{l}
R \\
J \end{array} \right \rangle {\hat{\cal T}}_{J^1}^{R^1}(1) {\hat {\cal 
T}}_{J^2}^{R^2}(2),
\end{equation}
where ${\hat {\cal T}}_J^R$ is the tensor operator of the coupling system. 
In order to calculate the matrix elements, we must choose the coupling 
scheme of the subsystem in the system. Suppose that we want to calculate 
the operator matrix elements between the left vectors belonging to the 
irreducible representation of $R^{1 \prime}$ and $R^{2 \prime}$ coupling 
into $R^\prime$ and the right vectors doing that of $R^{1 \prime \prime}$ 
and $R^{2 \prime \prime}$ coupling into $R^{\prime \prime}$. 

We can first make use the tensor product relation (34) and the 
Wigner-Eckart theorem (28) to obtain an expression for the reduced matrix 
element in such coupling scheme. The result is
\begin{equation}
\langle R^\prime \| {\hat {\cal T}}^R \| R^{\prime \prime} \rangle = 
[D(R^{\prime \prime})]^{-1} \left \{ \begin{array}{lll}
R^\prime & R & R^{\prime \prime} \\
R^{1 \prime} & R^1 & R^{1 \prime \prime} \\
R^{2 \prime} & R^2 & R^{2 \prime \prime} \end{array} \right \} \langle 
R^{1 \prime} \| {\hat {\cal T}}^{R^1}(1) \| R^{1 \prime \prime} \rangle 
\langle R^{2 \prime} \| {\hat {\cal T}}^{R^2}(2) \| R^{2 \prime \prime} 
\rangle,
\end{equation}
where we have defined the 9-J symbol of the $SO(5)$ angular momentum, which is given by
\begin{eqnarray}
\left \{ \begin{array}{lll}
R^\prime  & R & R^{\prime \prime} \\
R^{1 \prime} & R^1 & R^{1 \prime \prime} \\
R^{2 \prime} & R^2 & R^{2 \prime \prime} \end{array} \right \} & = &
\sum_{all~of~J's} \left \langle \begin{array} {ll}
R^1 & R^2 \\
J^1 & J^2 \end{array} \right. \left | \begin{array}{l}
R \\
J \end{array} \right \rangle \left \langle \begin{array}{ll}
R^{1 \prime \prime} & R^{2 \prime \prime} \\
J^{1 \prime \prime} & J^{2 \prime \prime} \end{array} \right. \left | 
\begin{array}{l}
R^{\prime \prime} \\
J^{\prime \prime} \end{array} \right \rangle \left \langle 
\begin{array}{ll}
R^{1 \prime} & R^1 \\
J^{1 \prime} & J^1 \end{array} \right. \left | \begin{array}{l}
R^{1 \prime \prime} \\
J^{1 \prime \prime} \end{array} \right \rangle \nonumber \\
& \times & \left \langle \begin{array}{ll}
R^{2 \prime} & R^2 \\
J^{2 \prime} & J^2 \end{array} \right. \left | \begin{array}{l}
R^{2 \prime \prime} \\
J^{2 \prime \prime} \end{array} \right \rangle \left \langle 
\begin{array}{l}
R^{\prime \prime} \\
J^{\prime \prime} \end{array} \right. \left | \begin{array}{ll}
R^\prime & R \\
J^\prime & J \end{array} \right \rangle \left \langle \begin{array}{l}
R^\prime \\
J^\prime \end{array} \right. \left | \begin{array}{ll}
R^{1 \prime} & R^{2 \prime} \\
J^{1 \prime} & J^{2 \prime} \end{array} \right \rangle.
\end{eqnarray}

Then, we scale the tensor operator ${\hat {\cal T}}_J^R$ as
\begin{equation}
{\tilde {\hat {\cal T}}}_J^R = \frac{{\hat {\cal T}}_J^R}{\langle R^\prime 
\| {\hat {\cal T}}^R \| R^{\prime \prime} \rangle}.
\end{equation}

Finally, from the tensor product relation of operators we can read off a 
fundamental formula of the coupling system's operators constructed by 
means of the tensor product of the subsystem's operators, which is
\begin{equation}
{\tilde {\hat {\cal T}}}_J^R = \sum_{R^2, R^{2 \prime}, R^{2 \prime 
\prime}, J^1, J^2} [D(R^{2 \prime \prime}) D(R^{1 \prime \prime})]^{-1} 
\left \langle \begin{array}{ll}
R^1 & R^2 \\
J^1 & J^2 \end{array} \right. \left | \begin{array} {l}
R \\
J \end{array} \right \rangle \left \{ \begin{array} {lll}
R^1 & R^2 & R \\
R^{1 \prime} & R^{2 \prime} & R^\prime \\
R^{1 \prime \prime} & R^{2 \prime \prime} & R^{\prime \prime} \end{array} 
\right \} {\tilde {\hat {\cal T}}}_{J^1}^{R^1}(1) {\tilde {\hat {\cal 
T}}}_{J^2}^{R^2}(2).
\end{equation}

It should be pointed that in the procedure of our obtaining the above 
formula, we have used the orthogonality relation of the 9-J symbol of the 
$SO(5)$ angular momentum. In fact, the inverse relation of the above formula also is 
important since it can be regarded as the operator product expansion of 
two operators. In order to make the transfer form one to another among 
them become convenient, here we write out this orthogonality relation
\begin{equation}
\sum_{R, R^\prime, R^{\prime \prime}} [D(R^{\prime \prime})]^{-1} 
\left \{ \begin{array}{lll}
R^1 & R^2 & R \\
R^{1 \prime} & R^{2 \prime} & R^{\prime} \\
R^{1 \prime \prime} & R^{2 \prime \prime} & R^{\prime \prime} \end{array} 
\right \} \left \{ \begin{array} {lll}
R^\prime & R & R^{\prime \prime} \\
R^{1 \prime} & R^1 & R^{1 \prime \prime} \\
R^{2 \prime} & R^2 & R^{2 \prime \prime} \end{array} \right \} = 
D(R^{2 \prime \prime}) D(R^{1 \prime \prime}).
\end{equation}

Now, we return to our considering the 4-dimensional quantum Hall system. 
At special filling factors, the quantum ground state of the 4-dimensional 
quantum Hall effect is separated from all excited states by a finite 
energy gap. This quantum ground state of the system of many particles is 
build up by some few of the particles lying in their LLL states in the 
coupling manner corresponding to a special filling factor. This quantum 
ground state is the LLL state of the system of many particles. The lowest 
energy excitations are the collective excitations of the system of many 
particles. Such collective excitations also can be built up by the lowest 
energy excitations of these particles in the fixed coupling manner. In 
fact, the matrix operators given in the above section are related with the 
generators of the lowest energy excitations. The present goal is to 
establish a scheme that the generators of matrix algebra corresponding to 
the collective excitations are constructed from the generators of the 
lowest energy excitations of the single particle. 

Because of the existence 
of the finite energy gap, all matrix operators act in their LLL Hilbert 
spaces. Generally, for the system including an Yang's $SU(2)$ monopole, 
its LLL Hilbert space is described by the irreducible representation of 
$SO(5)$ $R = [r_1 = r_2, r_2]$. This implies that for our considering 
system all irreducible representations labelling the matrix elements of 
the operators should belong to such irreducible representations of $SO(5)$ 
as $R = [r_1 = r_2, r_2]$, e.g., $R^\prime, R^{1 \prime}, R^{2 \prime}, 
R^{\prime \prime},R^{1 \prime \prime}$ and $R^{2 \prime \prime}$ in the 
equation (38) should be this kind of the irreducible representations of 
$SO(5)$. Furthermore, we have that $R^{\prime}=R^{\prime\prime}$, $R^{1 
\prime} = R^{1 \prime \prime}$ and $R^{2 \prime} = R^{2 \prime \prime}$ 
because the operators are realized by the matrices. Now, we can establish 
a fusion scheme of the matrices of the coupling system from the matrices 
of the subsystems based on the fundamental formula (38), which is
\begin{equation}
{\tilde {\hat {\cal Y}}}_J^R = \sum_{R^2, J^1, J^2} [D(R^{2 \prime})D(R^{1 
\prime})]^{-1} \left \langle \begin{array}{ll}
R^1 & R^2 \\
J^1 & J^2 \end{array} \right. \left | \begin{array}{l}
R \\
J \end{array} \right \rangle \left \{ \begin{array}{lll}
R^1 & R^2 & R \\
R^{1 \prime} & R^{2 \prime} & R^\prime \\
R^{1 \prime} & R^{2 \prime} & R^\prime \end{array} \right \} {\tilde {\hat 
{\cal Y}}}_{J^1}^{R^1}(1){\tilde {\hat {\cal Y}}}_{J^2}^{R^2}(2),
\end{equation}
where $R^\prime = [r_1^\prime = r_2^\prime, r_2^\prime]$, $R^{1 \prime} = 
[r_1^{1 \prime} = r_2^{1 \prime}, r_2^{1 \prime}]$ and $R^{2 \prime} = 
[r_1^{2 \prime} = r_2^{2 \prime}, r_2^{2 \prime}]$.

By using of the generalized triangular relation of the $SO(5)$ coupling 
coefficients and the property of  the 9-J symbol of $SO(5)$, a fusion rule 
of this fusion scheme can be read off as following
\begin{eqnarray}
|r_2^1 - r_2^2| \leq r_2 \leq r_1 \leq r_1^1 + r_1^2, \quad 
0 \leq r_2 \leq r_1 \leq 2 r_2^\prime, \nonumber \\ 
|r_2^{1 \prime} - r_2^{2 \prime}| \leq r_2^\prime \leq r_1^\prime \leq 
r_2^{1 \prime} + r_2^{2 \prime}, ~~~~ 0 \leq r_2^2 \leq r_1^2 \leq 2 
r_2^{2 \prime}, \nonumber  \\  0 \leq r_2^1 \leq r_1^1 \leq 
2 r_2^{1 \prime}.
\end{eqnarray}

For the 4-dimensional quantum Hall system composed of $N$ particles, one 
can repeatedly use the fusion formula and its fusion rule for $N-1$ times 
to obtain the elements of  the matrix algebra of this system. However, 
the matrix operators obtained in the finals are certainly those of the 
fuzzy monopole harmonics of the system. That is, if the LLL Hilbert space 
is the space that corresponds to the irreducible representation of $SO(5)$ 
$Q = [q_1 = q_2, q_2]$, the dimension of this Hilbert space  is $D(Q)$ 
and the operators obtained in the finals are the $D(Q)\times  D(Q)$ 
matrices. These matrix operators consist of the complete set of 
irreducible representations of $SO(5)$ in $D(Q) \times  D(Q)$ matrices, 
and their number is $D(Q)\times  D(Q)$. The non-commutativity of these 
operators is described by the matrix algebra (30) of replacing 
$\frac{P}{2}$ with $Q$. This matrix algebra is universal for the 
4-dimensional quantum Hall fluids. Hence, the matrix algebra is the 
description of the coordinate non-commutativity of particle's moving on 
$S^{4}$ in the 4-dimensional quantum Hall system. Exactly, this matrix 
algebra should be viewed as the description of non-commutative geometry of 
the 4-dimensional quantum Hall droplet since these matrix operators act in 
the LLL Hilbert space of our considering system.

Although both the equations (30) and (40) describe the operator product's relations 
of the fuzzy monopole harmonics,  the significances of them are different. The former 
should be view as the matrix multiplicater relation of the fuzzy monopole harmonics of 
the same system, and the latter as the operator tensor product relation of the fuzzy 
monopole harmonics of the different subsystems. Because the 4-dimensional quantum Hall droplet 
is a system of many-body, they determine all operators, which act in the LLL Hilbert space of 
the 4-dimensinal quantum Hall droplet, to obey the rules. As our explaining in the section 3, 
the fuzzy monopole harmonics can be considered as the generatores of the general wave functions 
of single particle system. Hence, the matrix algebra (30) and the operator tensor product (40) 
can be used the construction of the wave functions of the LLL  and its collective excitations of 
the 4-dimensional quantum Hall system. We shall discuss this construction elsewhere.

\section{Summary and outlook}

Similar to particle's motion on a plane in a constant magnetic field, the 
existence of Yang's $SU(2)$ monopole in the 4-dimensional quantum Hall 
system make the coordinates of particle's moving on the four-sphere become 
non-commutative. The appearance of such monopole also results in the 
irreducible representations of $SO(5)$ belonging to the Hilbert space of 
the system to be truncated. The similar phenomenon occurs in the 
description of fuzzy two-sphere. This clues us to the description of 
non-commutative geometry of the 4-dimensional quantum Hall system. Here we 
found that the fuzzy $S^{4}$ describes the non-commutative geometry of the 
4-dimensional quantum Hall droplet. By determining the explicit forms of 
fuzzy monopole harmonics and their matrix algebra, we established the 
description of non-commutative geometry of the 4-dimensional quantum Hall 
droplet. 

The $SU(2)$ monopole harmonics with the isospin rotating frame can be 
interpreted as  the wave functions of a 4-dimensional symmetrical top 
under the rotation transformationof $SO(5)$. Based on this view, we given 
the explicit expression of coupling coefficients of the $SO(5)$ angular 
momentum. The theory of angular momentum of $SO(5)$ is surprisingly simple 
and excellent, which is very similar to that of $SO(3)$. Many relations, 
paralleled the relations appearing in the $SO(3)$ theory, can be 
explicitly written off in $SO(5)$'s that. The expression of coupling 
coefficients given here is essential to the theory of angular momentum of 
$SO(5)$. These results are useful for the physics of atoms and molecules 
with the higher symmetry, i.e., $SO(5)$ symmetry. 

Following the proof of equivalence of two-dimensional quantum Hall physics 
and non-commutative field theory given recently by Hellerman and Raamsdonk 
\cite{Hellerman}, we can clarify the physical implication of the fuzzy 
monopole harmonics here. The second-quantized field theory description of 
the quantum Hall fluid for various filling fractions should involve some 
non-commutative field theory. On the 2-dimensional plane, such 
non-commutative field theory is  the regularized matrix version of the 
non-commutative $U(1)$ Chern-Simons theory. Our constructing the fuzzy 
monopole harmonics are a complete set of matrix version of non-commutative 
field theory corresponding to the 4-dimensional quantum Hall fluid. 
Because the creation and annihilation operators in the second-quantized 
field theory description of the quantum Hall fluid can be built up by this 
complete set. In this sense, these fuzzy monopole harmonics can be viewed 
as the creation and annihilation operators in the second-quantized field 
theory description of the quantum Hall fluid. The matrix algebra obeyed by 
these fuzzy monopole harmonics can be interpreted as the non-commutative 
relations satisfied by the creation and annihilation operators. The 
construction of the fuzzy monopole harmonics and their matrix algebra 
given by us is only the first step to establish the second-quantized field 
theory description of the 4-dimensional quantum Hall fluid, but it is an 
essential step. The fusion scheme of the fuzzy monopole harmonics and its 
inverse relation provide an approach for the calculation of correlation 
functions in the non-commutative field theory. In fact, the methods and 
some results present here can be straightforwardly used for the study of 
the 4-dimensional quantum Hall system.

On the other hand, it is interesting to relate the matrix algebra of fuzzy 
$S^{4}$ given here with some applications in D-brane dynamics in string 
theory and M-theory. The ability to construct the higher dimensional brane 
configurations using D0-branes is essential for a success of the Matrix 
theory of BFSS \cite{Banks}, where for example arbitrary membrane 
configurations in M-theory must be described in terms of the low energy 
degrees of freedom of the D0-branes.  Myers \cite{Myers} found that 
D0-branes expand into spherical D2-branes in the constant background RR 
fields. From the matrix model construction of Kabat and Taylor 
\cite{Kabat}, the non-commutative solution for such spherical D2-brane 
actually represents the bound state of a spherical D2-brane with some 
D0-branes. The fuzzy $S^{4}$ was used in the context of the Matrix theory 
of BFSS to describe time-dependent 4-brane solutions constructed from the 
D0-brane degrees of freedom. In this sense, the fuzzy monopole harmonics 
${\tilde {\hat {\cal Y}}}_J^R$ can be used to describe classical solutions 
of the the corresponding matrix brane model. However, the matrix algebra 
of fuzzy $S^{4}$ present here is different with those of fuzzy $S^{4}$ in 
the references \cite{Castellino,Constable,Ramgoolam,Ho}. The study of the 
matrix brane construction associated with the matrix algebra of fuzzy 
$S^{4}$ present here and the relation between it and the matrix brane 
constructions of the above references is an interesting topic. The work of 
this aspect is in progress.\\

We would like to thank K.J. Shi and L. Zhao for many valuable discussions. 
Y.X. Chen thanks the Institute of Modern Physics of Northwestern 
University in Xi'an for hospitality during his staying in the Institute. 
The work was partly supported by the NNSF of China and by the Foundation 
of Ph.D's program of Education Ministry of China.


\end{document}